\documentclass[10pt,letterpaper]{article}
\usepackage[top=0.85in,left=0.75in,footskip=0.75in,marginparwidth=2in]{geometry}

\usepackage[utf8]{inputenc}

\usepackage{cite}

\usepackage{nameref,hyperref}

\usepackage[right]{lineno}

\usepackage{microtype}
\DisableLigatures[f]{encoding = *, family = * }

\raggedright
\usepackage{changepage}

\usepackage[aboveskip=1pt,labelfont=bf,labelsep=period,singlelinecheck=off]{caption}

\makeatletter
\renewcommand{\@biblabel}[1]{\quad#1.}
\makeatother

\usepackage{lastpage,fancyhdr,graphicx}
\usepackage{epstopdf}
\fancyheadoffset[L]{2.25in}
\fancyfootoffset[L]{2.25in}

\usepackage{color}

\definecolor{Gray}{gray}{.25}

\usepackage{graphicx}

\usepackage{sidecap}

\usepackage{wrapfig}
\usepackage[pscoord]{eso-pic}
\usepackage[fulladjust]{marginnote}
\reversemarginpar

\begin{document}
\vspace*{0.35in}

% title goes here:
\begin{flushleft}
{\Large
\textbf\newline{Universal route to optimal few- to single-cycle pulse generation in hollow-core fiber compressors.}
}
\newline
% authors go here:
\\
E. Conejero Jarque\textsuperscript{1,*},
J. San Roman\textsuperscript{1},
F. Silva\textsuperscript{2,3},
R. Romero\textsuperscript{2,3},
W. Holgado\textsuperscript{1,+},
M.A. Gonzalez-Galicia\textsuperscript{1},
B. Alonso\textsuperscript{1},
I.J. Sola\textsuperscript{1},
H. Crespo\textsuperscript{2,1}
\\
\bigskip
\bf{1} Grupo de Investigaci\'on en Aplicaciones del L\'aser y Fot\'onica, Departamento de F\'isica Aplicada, University of Salamanca, E-37008, Salamanca, Spain.

\bf{2} IFIMUP-IN and Departamento de F\'isica e Astronomia, Faculdade de Ci\^{e}ncias, Universidade do Porto, R. do Campo Alegre 687, 4169-007 Porto, Portugal.

\bf{3} Sphere Ultrafast Photonics, S.A., R. do Campo Alegre 1021, Edif\'icio FC6, 4169-007 Porto, Portugal.

\bf{+} Present address: Centro de Laseres Pulsados, CLPU, Parque Cient\'ifico, Villamayor, Spain.

\bigskip
* enrikecj@usal.es

\end{flushleft}

\section*{Abstract}
Gas-filled hollow-core fiber (HCF) pulse post-compressors generating few- to single-cycle pulses are a key enabling tool for attosecond science and ultrafast spectroscopy. Achieving optimum performance in this regime can be extremely challenging due to the ultra-broad bandwidth of the pulses and the need of an adequate temporal diagnostic. These difficulties have hindered the full exploitation of HCF post-compressors, namely the generation of stable and high-quality near-Fourier-transform-limited pulses. Here we show that, independently of conditions such as the type of gas or the laser system used, there is a universal route to obtain the shortest stable output pulse down to the single-cycle regime. Numerical simulations and experimental measurements performed with the dispersion-scan technique reveal that, in quite general conditions, post-compressed pulses exhibit a residual third-order dispersion intrinsic to optimum nonlinear propagation within the fiber, in agreement with measurements independently performed in several laboratories around the world. The understanding of this effect and its adequate correction, e.g. using simple transparent optical media, enables achieving high-quality post-compressed pulses with only minor changes in existing setups. These optimized sources have impact in many fields of science and technology and should enable new and exciting applications in the few- to single-cycle pulse regime.

% now start line numbers
%\linenumbers

% the * after section prevents numbering
\section*{Introduction}
The chirped pulse amplification (CPA) technique applied to Titanium Sapphire lasers has made intense near-infrared (NIR) ultrashort pulses in the $20-100$ fs range widely available for scientific, biomedical and industrial applications. Special efforts have been devoted to generate even shorter pulses, in the few- and single-cycle regime, due to their potential applications and interest. In particular, such pulses have paved the way for attosecond physics and metrology \cite{hentschel01,kling2008,krausz2009,gallmann2012,krausz2014,timmers2017}, via the extreme ultraviolet (XUV) attosecond pulse trains and isolated attosecond pulses \cite{sansone06} that can be obtained by high-harmonic generation (HHG). The use of few-cycle optical pulses with durations close to or shorter than 10 fs in the near-infrared, visible and near-ultraviolet spectral regions has been extended in recent years to a wide range of spectroscopic techniques such as impulsive vibrational spectroscopy \cite{polli2008,liebel2015,du2016}, time-resolved stimulated Raman spectroscopy \cite{kukura2007,fujisawa2016,dietze2016,kuramochi2016}, and ultrafast pump-probe absorption spectroscopy \cite{wang2015,luo2016}. Few-cycle optical pulses have also become an interesting tool for transient absorption microscopy \cite{schnedermann2016}, near-field imaging techniques \cite{nishiyama2015} and for generating ultrashort terahertz radiation \cite{darmo2004}.

While it is possible to obtain sub-10 fs pulses from CPA \cite{seres03} or from optical parametric amplification \cite{cerullo98} systems, the former is not easy to accomplish, and the latter is not commonplace. Therefore, post-compression techniques are usually employed for the generation of intense few- and even single-cycle pulses in the near- and mid-infrared spectral regions\cite{miranda12b,cardin15}. In order to post-compress ultrafast pulses down to the few-cycle regime, two steps are usually needed. First, nonlinear processes broaden the pulse spectrum, thus decreasing the Fourier limited pulse duration. In a second step, the spectral phase resulting from the previous stage is compensated, typically using chirped mirrors, gratings, prisms, or other dispersive systems, resulting in a temporally compressed pulse. This scheme was first proposed in the context of optical fibers in the 1980s \cite{tomlinson84}, and enabled achieving 6 fs pulses when compensating simultaneously the outcoming group delay dispersion (GDD) and third-order dispersion (TOD) \cite{fork87}. The scheme was successfully expanded in 1996 by Nisoli and coworkers to the ultra-intense laser pulse regime (mJ-level pulses) by using hollow-core fibers (HCF) filled with gases \cite{nisoli96}. Using the latter scheme, together with chirped mirrors as the compression system, few-cycle pulses in the hundreds of $\mu$J energy range with 0.1 TW peak power were obtained \cite{sartania97}. Later, similar results were obtained using the light filamentation process in the spectral broadening stage \cite{hauri04}. 

In spite of requiring a finer input beam alignment than filamentation-based compressors, HCF compressors are today the most widely used high-energy post-compressed sources, in part due to their intrinsic spatial filtering properties which result in very high quality beam profiles. Furthermore, their long-term stability can be greatly improved by using, e.g., piezo-driven mirror mounts to ensure stable and constant spatial coupling of the input laser pulses into the HCF via a simple feedback loop. Hollow-core fiber post-compressed pulses have shown a great potential in a wide range of applications, such as pump-probe experiments in conjunction with attosecond pulses \cite{blattermann2015}, ultrafast measurement of electrical and optical properties of solids \cite{schiffrin2013,schultze2013}, time-resolved studies of Coulomb explosion dynamics \cite{bocharova2011}, ultrafast spectroscopy techniques \cite{liu2010,liu2013,kobayashi2012,gueye2016,paolino2016,goncalves2016,chang16}, and very recently a new generation of compact kHz laser-plasma accelerators based on single-cycle pulses \cite{guenot16}.

To access the Fourier Limit of a pulse after a nonlinear propagation process one has to deal with the complex phase that the pulse acquires due to the interplay of different linear and nonlinear effects. In general, researchers optimize their HCF compressors by empirically adjusting several key parameters, such as gas type and pressure, input pulse characteristics and coupling conditions, with the final result usually involving a delicate compromise between output efficiency, amount of spectral broadening and achievable degree of compression (pulse duration and quality) for their particular system and chirped mirror set. The ability to measure and quantify the achieved degree of compression is therefore paramount to identify the main characteristics of the output pulse and to further optimize its compression. Both spectral phase oscillations and the overall spectral phase of a pulse can be visualized in a very straightforward way using the dispersion-scan (d-scan)\cite{miranda12} technique, which has been extensively used in the last years to characterize many state-of-the-art few-cycle pulse sources around the world and is enabling new and very promising applications \cite{goncalves2016,louisy15,chang16,guenot16,timmers2017}.

D-scan is a recent approach for the simultaneous measurement and compression of femtosecond laser pulses. Its experimental setup is fully in-line and does not require beam splitting and recombination nor temporal delay of short pulses. Experimentally it involves the measurement of the spectrum of a nonlinear signal such as second-harmonic generation (SHG) as a function of dispersion applied to the pulse. This can be performed with pulse compression setups, such as a chirped mirror (CM) and glass wedge compressor, where the amount of glass traversed by the pulse is an independent variable controlled by insertion of one of the wedges: while the CMs impart negative dispersion, the variable positive dispersion introduced by the wedges will vary the total dispersion experienced by the pulse to be measured. In second-harmonic generation  d-scan (SHG d-scan), measuring the SHG signal after the compressor provides a two-dimensional trace of the SHG spectrum vs. insertion. An optimization algorithm is then used to retrieve the spectral phase of the pulse from the measured d-scan trace and calibrated linear spectrum \cite{miranda12}. A recent approach to d-scan retrieval can also be used to obtain the pulse amplitude and phase from the measured trace \cite{miranda16}, but in this case, the trace itself must be calibrated. 

D-scan has been successfully demonstrated with few-cycle pulses since its inception \cite{miranda12,miranda12b}, and over octave-spanning single-cycle pulses have been measured directly with SHG d-scan \cite{silva14,fabris15,miranda16,guenot16,timmers2017}. Apart from its robustness and performance, another important advantage of d-scan is the fact that it directly results in very intuitive traces that provide useful guiding information on the quality of the achieved pulse compression, which motivates our use of the d-scan trace as a diagnostic tool. For instance, a flat and thin trace is indicative of excellent compression, since this means that for a particular position of the compressor, all spectral components are equally compressed and hence their SHG signal is maximized. If the trace has a tilt, this means that different parts of the spectrum are being compressed for different dispersions. In other words, we have a frequency-dependent chirp in the pulses, i.e., the pulses have third-order dispersion. A curved, parabolic-like trace would be indicative of fourth-order dispersion, and so on. In the case of phase oscillations, these will produce spectral modulations and a wavy appearance in the resulting d-scan trace (for a more detailed description of d-scan traces and their interpretation, the reader is referred to refs. \cite{miranda12,miranda12b}).

In this paper we look for a route to obtain the optimum stable post-compressed pulse from a HCF using the d-scan technique as the compression device and the d-scan trace to univocally identify this route. Numerical simulations and experimental measurements reveal that, in quite general conditions, the best performance results in the post-compressed pulses typically exhibiting a residual third-order dispersion of nonlinear origin, i.e., a signature TOD which is intrinsic to optimum nonlinear propagation within the fiber, in agreement with experimental observations made in several laboratories around the world. Overdriving the HCF above this optimum regime invariably results in an increasingly complex nonlinear spectral phase which renders compression very hard to optimize. These results and behavior have been obtained for different gases and different setup parameters (including HCF length, gas pressure and pulse energy) showing the universality of the existence of the optimum regime. We have also probed the spatial quality of the optimum output pulse, showing a high spatio-spectral homogeneity. Moreover, we have been able to develop a simple theoretical model that explains the main nonlinear effects underlying the optimum regime, which is very useful for finding out the proper parameters to achieve the desired few- to single-cycle pulses. The understanding of the optimum propagation regime, of its intrinsic TOD and its subsequent correction using, e.g., unusual transparent optical media with adequate ratio between second- and third-order dispersion, enables achieving optimized high-quality post-compressed pulses with only minor changes in existing setups.

\section*{Results and Discussion}

\subsection*{Regimes of nonlinear propagation in the hollow-core fiber. Identification of the optimum d-scan trace.}

The identification of the optimum parameters of a nonlinear process to obtain a desired output pulse is not an easy task. Fortunately, in the context of nonlinear propagation of ultrashort laser pulses, we have helpful theoretical models to guide us. We used a nonlinear spatio-temporal model (see the Methods section) to simulate the nonlinear propagation of laser pulses in a static HCF filled with different gases (argon, neon and air) and in different conditions (gas pressure; input pulse energy and duration), and their subsequent compression with a d-scan system based on chirped mirrors and glass wedges. All cases studied are representative of real experimental situations and present qualitatively similar dynamics.

\begin{figure}[htbp]
\centering\includegraphics[width=14cm]{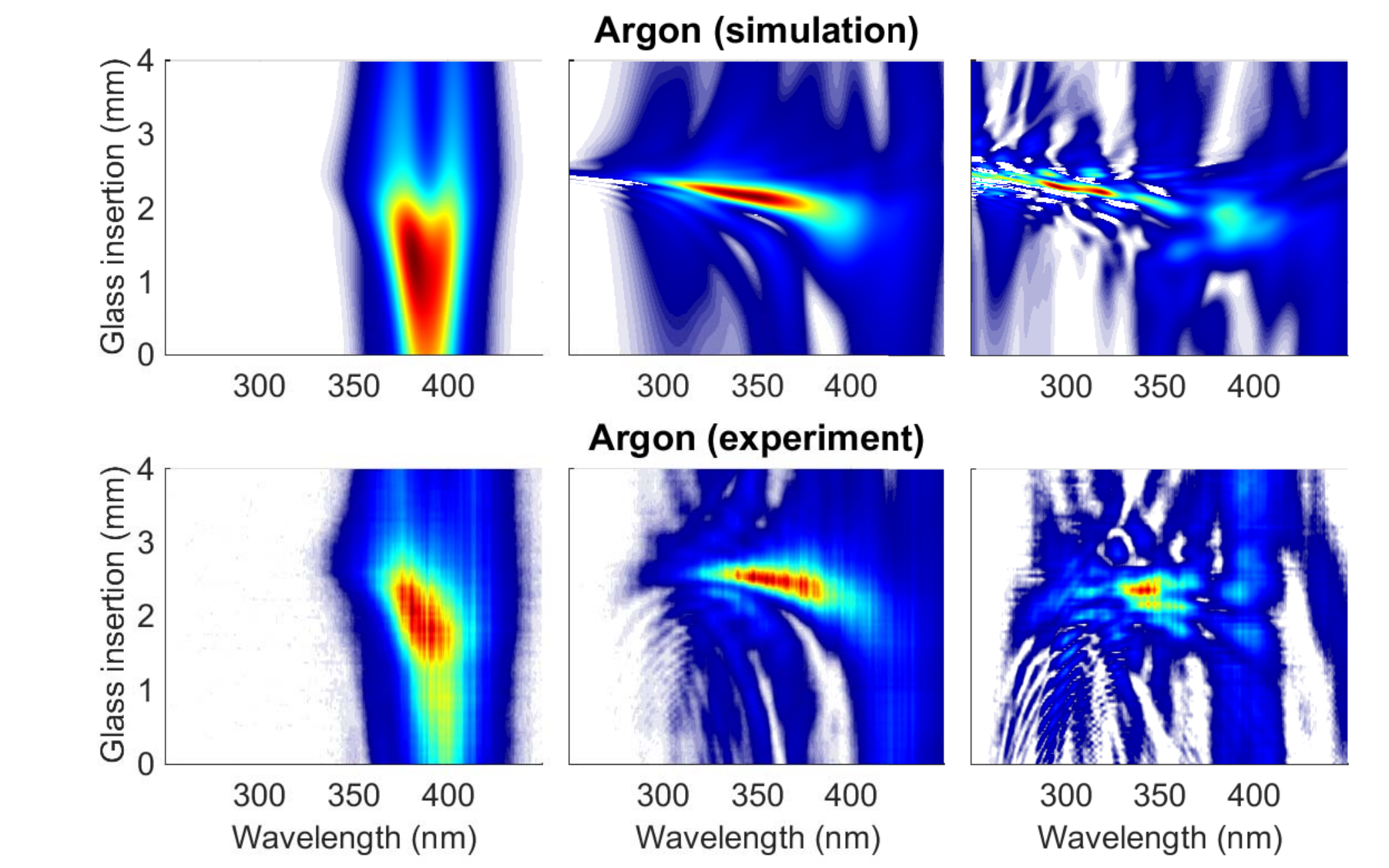}
\caption{The theoretical d-scan traces (top row) correspond to the propagation of a Fourier limited 25 fs full-width at half-maximum (FWHM) Gaussian pulse, with 0.2 mJ of energy and with the spectrum centered at 780 nm, in a 25 cm long HCF with a core diameter of 250 micron and filled of Ar at 0.2 bar (left), 0.6 bar (middle) and 1.0 bar (right). All the simulations are done assuming that the beam couples perfectly into the fundamental mode of the HCF. The experimental d-scan traces (bottom row) correspond to the propagation of $\sim$\,25 fs FWHM pulses with 1 mJ energy centered at 780 nm in a 1.0 meter long HCF with a core diameter of 250 micron filled with argon at 0.15 bar (left), 0.265 bar (middle) and 0.9 bar (right).}
\label{fig1}
\end{figure}

In Fig.\,\ref{fig1} we present the theoretical (top row) and experimental (bottom row) d-scan traces obtained in a pressure-scan experiment. The left, middle and right column represent the low pressure cases (low interaction regime), the optimum pressure cases (optimum interaction regime) and the high pressure cases (high interaction regime), respectively (see the figure's caption for the detailed set of parameters). The theoretical d-scan traces are calculated from the on-axis field obtained at the end of the HCF. The d-scan compressor is composed of chirped mirrors followed by a pair of BK7 glass wedges, as in the d-scan setup commonly used for few-cycle pulse compression \cite{miranda12}, but taking only into account their group delay dispersion effect and neglecting higher order terms. We assume ultra-broadband chirped mirrors, which introduce -120\,fs$^2$ of pure GDD, with the BK7 wedges introducing 46.6\,fs$^2$/mm at the central wavelength 780 nm. This simplification helps us identify the origin of the higher order dispersion terms that may appear in the pulse phase, which must come from the propagation in the HCF because the compression/measurement stage does not introduce them. Therefore, and based on the explanation given in the Introduction section, we will be able to identify any high order phase terms of the output pulse directly from the d-scan trace structure.

The observed similarity between the theoretical and experimental pressure scans indicates that the theoretical model, including the calculation of d-scan traces, takes into account the main effects occurring in the experiment. Regarding the calculation of the theoretical d-scan traces we should note that real CMs introduce, by design, a given amount of negative TOD to compensate for the positive TOD of the gas, window and wedge materials used in the post-compression stage. In view of the good agreement between the theoretical and the experimental d-scan traces shown in Fig.\,\ref{fig1}, the TOD that we have neglected in our theoretical compression/measurement stage (CMs and wedges) is indeed a good approximation of an actual experiment.

The results presented in Fig.\,\ref{fig1} also help us identify the optimum d-scan traces (those in the middle column). It is clear that the low interaction regime traces (left column) represent situations where the pulses did not have enough nonlinear interaction to broaden their spectra, showing narrow d-scan traces in the spectral coordinate. In contrast, the strong modulations of the high interaction regime traces (right column) indicate that they correspond to situations where the pulses had such intense nonlinear interaction that, although the spectral broadening was large, the complex spectral phase makes the output pulse useless for most applications. Moreover, this regime is easily identified in the experiments due to a poor output stability. We can therefore define the optimum interaction regime as the one that generates d-scan traces similar to those of the middle column, which represent situations where the pulses are spectrally broadened but in a regime in which the acquired spectral phase can be adequately compensated for. The retrieved pulses obtained from the optimal d-scan traces presented in Fig.\,\ref{fig1} have durations of $\sim$ 4.1 fs FWHM for the theoretical case (top row) and $\sim$ 4.0 fs FWHM for the experimental case (bottom row). These optimum d-scan traces already correspond to sub-two-cycle pulses, and additional compensation of their intrinsic residual TOD \cite{silva14} allows obtaining high-quality pulses in the single-cycle regime\cite{silva14,miranda16,timmers2017}.

The exact limits of the three identified propagation regimes are somewhat flexible but, as shown in Fig.\,\ref{fig1}, they can be easily identified using the d-scan as the guiding tool. There is a large number of parameters that one can use to reach the optimum regime in the laboratory: input pulse energy, input pulse temporal duration (chirp), coupling conditions (focusing, mode and numerical aperture matching), gas type and pressure. Using some of them one can get into the optimum regime to achieve {\it the optimal post-compressed output pulse, which will not be the one with the broadest spectrum, but the shortest among the less temporally-structured obtained pulses}, which we have been able to univocally identify through the resulting d-scan trace. 

\subsection*{Optimum d-scan trace properties: Universality, Spatial and Spectral structure, and Physical origin.}
 
\begin{figure}[htbp]
\centering\includegraphics[width=14cm]{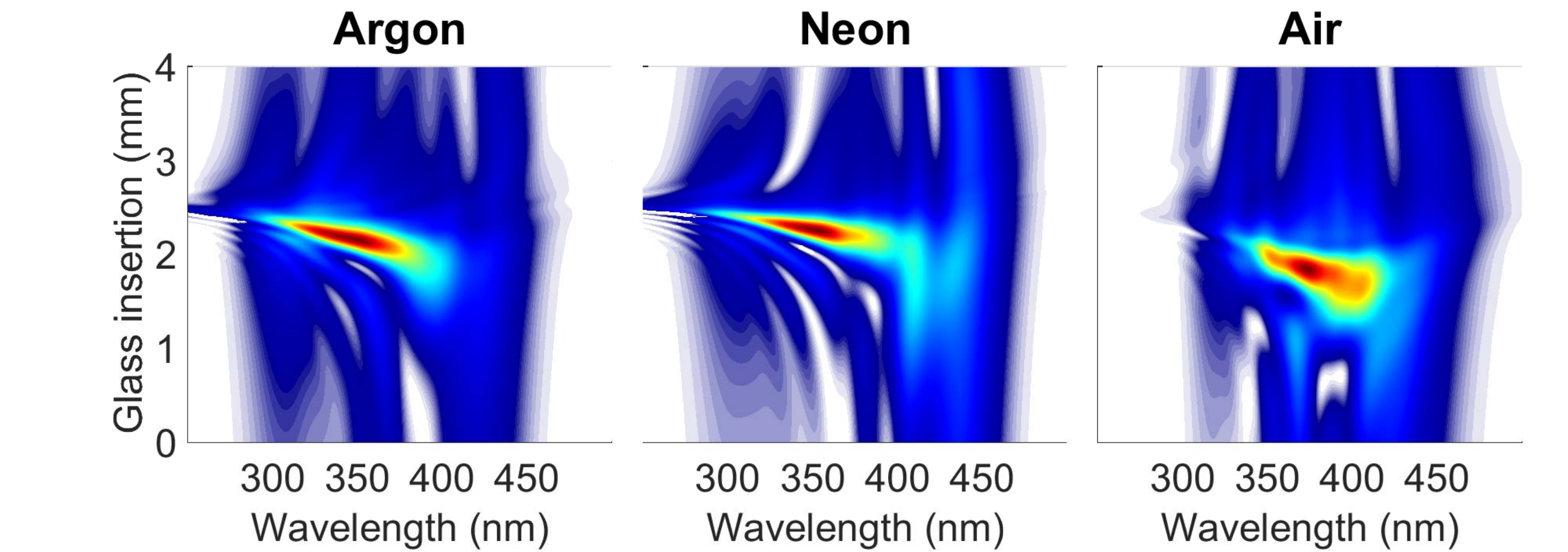}
\caption{Simulated d-scan traces for a pulse with 0.2 mJ energy propagating in a 25 cm long HCF filled with argon at 0.6\,bar (left), a pulse with 0.4 mJ propagating in a 50 cm long HCF filled with neon at 2.0\,bar (middle), and a pulse with 0.15 mJ propagating in a 75 cm long HCF filled with air at 0.5\,bar (right). All the simulations assume a perfect coupling into the fundamental mode of a HCF with a core diameter of 250 micron, and an input pulse with 25 fs FWHM, spectrally centered at 780 nm.}
\label{fig2}
\end{figure}

In the previous results, we have been able to identify the optimum d-scan trace when using argon. We now apply the theoretical model to verify that similar optimum d-scan traces appear using different gases and experimental parameters. Figure \ref{fig2} shows the theoretical optimum d-scan traces obtained at the end of a HCF using argon (left), neon (middle) and air (right). The retrieved output pulse duration for the three cases is $\sim$ 4.1 fs, $\sim$ 4.0 fs and $\sim$ 4.7 fs FWHM, respectively. All these traces were identified after simulating pressure scans as the one presented in the top row of Fig.\,\ref{fig1}. When using argon or neon we obtained almost identical optimum traces, while when using air there are some variations due to the presence of the Raman effect, which is not present when using atomic gases. Surprisingly, regardless of the different conditions (fiber length, pulse energy, gas pressure and gas type), we have always been able to identify the optimum d-scan trace at the end of the HCF. This universal behavior, together with the particular structure of the optimum d-scan trace as a univocal fingerprint, is a very useful tool both for optimizing existing systems and to promote the spreading of few-cycle pulse systems for many more applications that can benefit from a stable and reproducible source.

For completeness we present a movie (see Supplementary Material) showing the evolution of the d-scan trace at the end of the HCF during a pressure scan with argon, together with the evolution of the wedge-insertion-dependent post-compressed pulse. The left plot in the movie shows the particular features of the d-scan trace along all three stages. The right plot shows the final compressed pulse that can be obtained, displaying the continuous evolution of the structure of the output pulse.

The structure of the optimum d-scan trace gives us important information about the output pulse obtained from the post-compression system. The first distinctive feature of the optimum d-scan trace is a slightly negative slope which corresponds to a negative TOD. This is the remaining residual TOD accumulated during the nonlinear propagation of the pulse inside the HCF. According to the simulations, this TOD mostly comes from the self-steepening effect, similarly to what occurs in post-compression setups for pulses in the mid-infrared spectral region \cite{fan16}. This tilt of the d-scan trace is very helpful to find experimentally the optimum regime to obtain single-cycle post-compressed pulses. To our knowledge, this remaining TOD is clearly present in experimental HCF post-compressed pulses obtained in different laboratories around the world when optimizing for the output pulses (e.g., United Kingdom\cite{fabris15}, Portugal\cite{silva14,fabris15,Alonso13}, France\cite{bohle14}, Sweden\cite{louisy15,heyl16}, Germany\cite{tajalli16}, Canada\cite{schmidt10}, China\cite{huang16}, Austria\cite{fan16} and United States of America\cite{chang16,timmers2017}, as some examples). Those experiments were done using HCFs with different characteristics (length and/or core diameter), filled with different gases (Ar, Ne or He) and with lasers in the NIR and mid-IR regions. All these observations support that the route to obtain an optimum post-compressed pulse described here is universal.

\begin{figure}[htbp]
\centering\includegraphics[width=17cm]{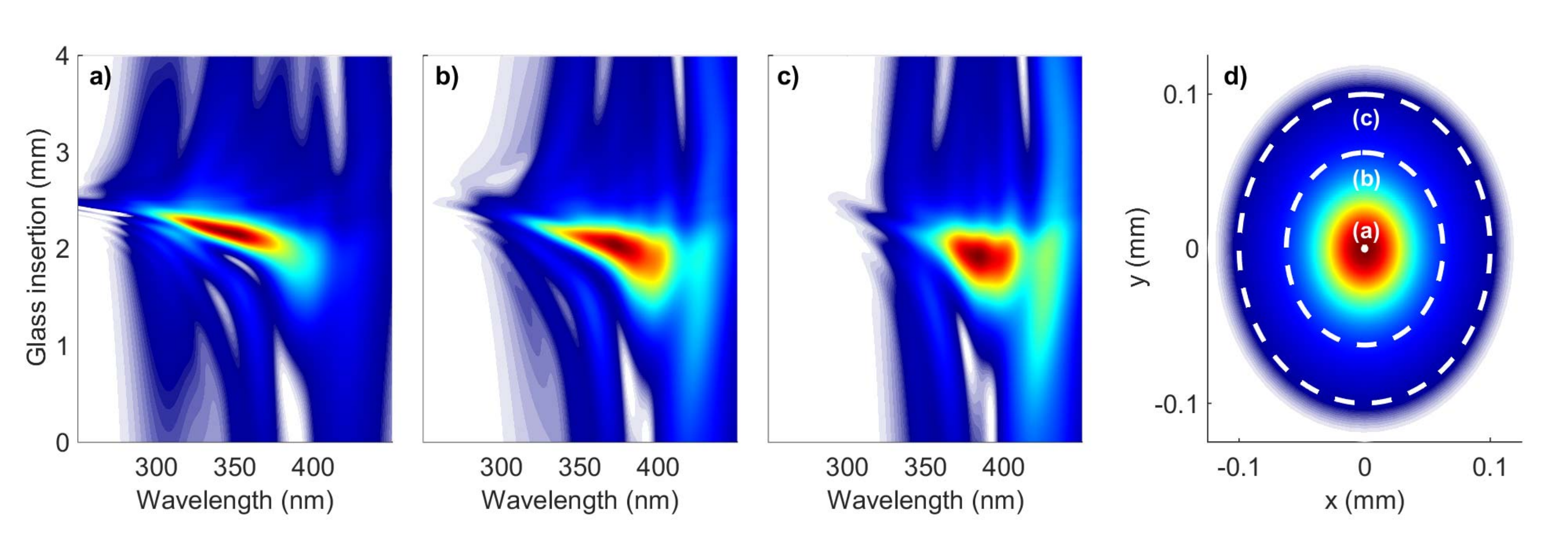}
\caption{Simulated far-field d-scan traces obtained by integrating the beam with an on-axis aperture with a radius of 1 $\mu$m (taking only into account the on-axis field) (a); 62.5 $\mu$m (b); and 100 $\mu$m (c). All the d-scan traces correspond to the case of a 25 fs FWHM input pulse, spectrally centered at 780, with 0.2 mJ of energy, propagating in a 25 cm long HCF filled with argon at 0.6\,bar. Like before, a perfect coupling into the fundamental mode of the HCF is assumed. In (d) the fluence of the beam at the end of the HCF is shown, with circles indicating the different apertures used to calculate the three far-field d-scans.}
\label{fig3}
\end{figure}

Another very important feature of the output pulse related to the optimum d-scan is spatial homogeneity. All the theoretical d-scan traces presented until now have been calculated from the on-axis field at the end of the HCF. One would expect that, as the optimum regime corresponds to a moderate nonlinear interaction regime, the resulting output beam should have a quite good spatial homogeneity, as it has in fact been proved experimentally in similar conditions \cite{Alonso13}. Figure \ref{fig3} shows the theoretical far-field d-scan calculated from a spatial selection of the field obtained at the end of the HCF. The parameters here are the same used to obtain the optimum d-scan trace shown in the leftmost case of Fig. \ref{fig2}. The far-field distribution is calculated by doing the Hankel transform integrating spatially the selected part of the beam. Then the d-scan is finally calculated using the far-field corresponding to $k_{\perp} =0$, i.e. the zero divergence far-field. Figure \ref{fig3}a) shows the d-scan trace of the far-field when taking only into account the on-axis near field, which is basically the same presented using the field on-axis (left picture of Fig.\,\ref{fig2}). Figure \ref{fig3}b) shows the d-scan trace of the far-field when integrating the near-field up to 62.5 $\mu$m (half of the HCF core radius, which contains 94\% of the output energy). As it can be observed, the d-scan trace essentially retains the same shape as before, being slightly narrower in the spectral direction and more stretched along the dispersion axis. Figure \ref{fig3}c) shows the d-scan of the far-field obtained by integrating up to 100 $\mu$m (which accounts for 99\% of the output energy). In this case the spectral narrowing of the d-scan is more pronounced, as we are taking into account the most external part of the beam that sees less nonlinearity. Note that even this last d-scan trace preserves the TOD signature related to the optimum regime, although the optimal pulse duration changes from $\sim$ 4.1 fs in the first case to $\sim$ 4.7 fs and $\sim$ 5.3 fs in the second and third cases, respectively. Finally Fig.\,\ref{fig3}d) shows the beam fluence at the end of the HCF with circles indicating the integrated area used to calculate the far-field d-scan for the three cases shown, which helps to visualize the good homogeneity of the beam obtained under the optimum parameters. The smooth output spatial profile also shows that the post-compressed pulse is suitable for applications which require good beam quality.

The identification of the optimum regime and the spatial homogeneity of the obtained output beam indicate that the optimum regime is achieved under a moderate nonlinear interaction regime. To unveil the main nonlinear effects underlying this regime, we have looked for a simple theoretical model that could explain the results. This model could be very useful to find out the proper parameters to reach the optimum regime and, as a consequence, to achieve the desired few- or single-cycle post-compressed pulses. The model assumes that the self-phase modulation (SPM) is the main nonlinear effect responsible for the output pulse structure. We have then estimated the B-integral accumulated during the propagation of a pulse with energy $E_{in}$, temporal duration $t_p$, and spectrum centered at $\lambda_0$, propagating in a HCF of length $L_F$ and with a core of radius $r_F$. We assumed that the beam couples perfectly to the fundamental mode of the HCF, with well-known losses denoted by $\alpha$ \cite{marcatili64}. Inside the HCF we have a gas at constant pressure and with nonlinear index $(1-x_R )n_2$, where $x_R$ only indicates the amount of Raman effect ($x_R =0$ for atomic gases and $x_R \sim 0.6$ for molecular gases like air). The accumulated B-integral then reads:
\begin{equation}
B=\frac{2\pi}{\lambda_0} \int_0^{L_F} n_2 I (z) dz = \frac{2\pi}{\lambda_0} I_0 (1-x_R )n_2 \frac{\left( 1-\exp \left( -\alpha L_F \right) \right)}{\alpha}.
\label{eq_bint}
\end{equation}
To estimate the input intensity we use the following simple formula: $I_0 = E_{in}/(t_p \pi r_F^2 )$. Using these expressions we have verified that for all the simulations in which we have observed the optimum d-scan trace at the end of the HCF, we obtain a similar value of the B-integral, $B^{th} > 4$ (e.g., we obtained values of 4.2, 4.4 and 5.6 for the three situations presented in Fig.\,\ref{fig1}, corresponding to argon, neon and air, respectively). A lower limit value of $B_{opt}^{th} =4$ gives us a very useful tool to find out the good parameters to achieve the optimum regime. Taking into account that the nonlinearity depends linearly on the gas pressure, we have used Eq. \ref{eq_bint}  to calculate, for example, the pressure needed to obtain the optimum d-scan trace at the end of a HCF of fixed length and core radius ($L_F$,$r_F$) filled with some gas with nonlinear refractive index $(1-x_R )n_2^{*}$ (being $n_2^{*}$ the nonlinear parameter at 1 bar pressure), with an input pulse of energy $E_{in}$ and a temporal duration $t_p$:
\begin{equation}
P_{opt}^{th} =B_{opt}^{th} \frac{\lambda_0 \alpha}{2\pi I_0 (1-x_R )n_2^{*} \left( 1-\exp \left( -\alpha L_F \right) \right)} .
\label{eq_popt}
\end{equation}
We have checked that using the parameters obtained from Eq. \ref{eq_popt} brings one to a situation very close to the optimal d-scan trace for argon, neon and air for a wide range of parameters, whenever the propagation is in a moderate nonlinear regime (not presenting, for instance, self-focusing dynamics or similar phenomena related to intense nonlinear interactions). These expressions demonstrate the tunability of the optimal trace with respect to key parameters such as pressure, type of gas and input energy or pulse duration, setting this procedure as a universal route to obtain few-cycle pulses in the visible-NIR spectral region.

We have estimated the optimum B-integral for some experimental cases\cite{Alonso13,timmers2017,chang16,silva14}, besides the experiments presented in this work, obtaining a value $B_{opt}^{exp} > 12$ (range of values from 12 to 23 with an average of 17). To calculate these values we have estimated that the energy coupled into the HCF is always half of the input energy available and we have assumed that the input pulse is Fourier limited. Even with all these assumptions, which are not always fulfilled in experiments, we have obtained a roughly constant experimental optimum B-integral, which could be very useful to bring an experiment very close to the optimum regime. As expected, $B_{opt}^{exp} > B_{opt}^{th}$ due to the more effective nonlinear interaction present in the numerical models than in real experiments.

\subsection*{TOD compensation of the optimal d-scan trace.}

The residual TOD which appears in the optimal output pulses can be further compensated for, not only to reduce the output pulse duration but also to increase the amount of energy in the output pulse. This can be done, for example, by using an adequate transparent medium such as water \cite{silva14,fabris15,louisy15}, z-cut KDP \cite{miranda16} or z-cut ADP \cite{timmers2017}, which directly enables achieving high-quality pulses in a more strict single-cycle regime. In spite of the relatively small magnitude of this TOD (e.g., -40 \,fs$^3$ reported in Ref. \cite{silva14}), its effect on a few- and single-cycle pulse can be dramatic: its compensation enabled going from a 3.8\,fs pulse to a near-transform-limited 3.2\,fs 1.3-cycle pulse, which was accompanied by a significant improvement in pulse contrast and peak intensity \cite{silva14}.

\begin{figure}[htbp]
\centering\includegraphics[width=14cm]{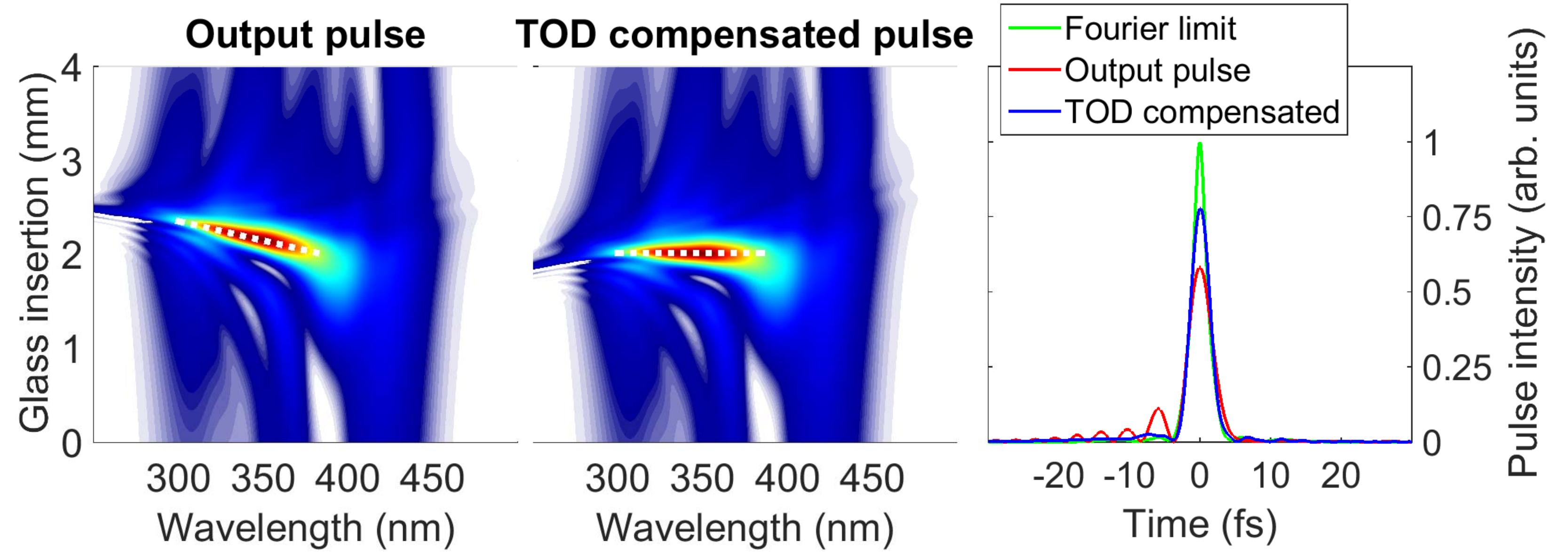}
\caption{D-scan traces of (left) the optimum pulse for the simulation parameters of the top middle graph in Figure \ref{fig2} (Ar-filled HCF with 0.6\,bar pumped with Fourier-limited $\sim$25 fs pulses) and (center) the same pulse with additional TOD compensation (+20\,fs$^3$). The dashed white lines are a visual guide to see the change of slope in the trace associated to TOD compensation. (right) Output pulse in the time domain, before (red line) and after (blue line) TOD compensation. The Fourier-limited pulse (green line) is also shown for reference.}
\label{fig4}
\end{figure}

In the center plot of figure \ref{fig4} we show the d-scan trace obtained after compensating the residual TOD by adding +20\,fs$^3$ at the central wavelength, 780 nm, in the measurement/compression stage, compared with the non-compensated case (left plot). We see that TOD compensation effectively tilts the d-scan trace so that its average negative slope practically disappears. To illustrate the improvement in intensity throughput after TOD compensation we show on the right plot of Fig.\,\ref{fig4} the direct output pulse and the TOD-compensated pulse, together with the Fourier-limited pulse. The used glass insertion in this case is the one which results in maximum intensity in the d-scan trace. Clearly, the pulse corresponding to the TOD-corrected trace has better features: it presents an increase in peak intensity of the order of 33\%, a reduction of the temporal FWHM pulse duration from $\sim$ 4.1 fs to $\sim$ 2.9 fs, which is in very good agreement with experimental results, and smaller secondary temporal structures than the non-corrected case, as also confirmed by experiments.

%\section*{Discussion}

%The Discussion should be succinct and must not contain subheadings.

\section*{Conclusion}

In this work we have presented and analyzed a route to obtain optimum hollow-core fiber post-compressed pulses using the d-scan setup as the compression and diagnostic device. We have demonstrated that by changing the gas pressure and/or input pulse energy, one is able to identify the optimal post-compression parameters to obtain the shortest, less structured and stable output pulses. This optimal setup can be univocally identified by the d-scan trace of the output pulses, which shows a marked TOD feature, whereas overdriving the HCF above this optimum regime invariably results in an increasingly complex nonlinear spectral phase, which renders compression very hard to optimize. We have verified that the optimum d-scan trace can be found for different gases and setup conditions, highlighting the universality of this phenomenon. Moreover, we have also proved the good spatio-spectral characteristics of the output pulse obtained under these conditions. For the optimum propagation region, the remaining TOD can be further corrected, for example by simple linear propagation in an adequate transparent medium, which improves the temporal shape and the peak power of the output pulse. The identified propagation regime and approach enable the generation of stable and high-quality few- to single-cycle pulses, which has direct implications in the performance of current and new HCF pulse post-compression systems and will help improve and extend the applications of these extreme light sources in many fields of science and technology.

\section*{Methods}

{\bf Simulations.} To study the nonlinear propagation of a laser pulse in a HCF we have implemented the standard nonlinear envelope propagation equation \cite{couairon07}. The model includes the spatial and temporal pulse dynamics. We use a local frame moving with the pulse $T=t-z/v_g$, being $v_g$ the group velocity of the pulse, and assume cylindrical symmetry ($r$ radial coordinate, $z$ axial coordinate). The propagation equation for the temporal envelope of the pulse, $A(r,z,T)$, is
\begin{equation}
\label{eq1}
\frac{\partial A(r,z,T)}{\partial z} = (\hat{L}+\hat{N})A(r,z,T).
\end{equation}
The first part of the propagation equation, $\hat L$, represents the linear propagation effects: diffraction, dispersion and linear losses. The other part, $\hat N$, represents the nonlinear propagation effects, which include self-phase modulation, Raman scattering, ionization, losses due to the ionization process and plasma absorption, and self-steepening.

To solve Eq. \ref{eq1} we use a split-step method dividing each propagation step into two sub-steps \cite{agrawal}. The first one consists on applying only the linear effects by decomposing the input pulse into the $EH_{1m}$ modes of the hollow-core fiber\cite{marcatili64,granados12}
\begin{equation}
\label{eq2}
\tilde A(r,z,\omega)=\sum_{m=1}^{\infty} {c_m (\omega,z) EH_{1m} (r)},
\end{equation}
where $\tilde A(r,z,\omega)$ represents the Fourier Transform of $A(r,z,T)$. The coefficients of the decomposition can be calculated by doing the inverse Hankel Transform of the spatial beam distribution in the core of the HCF
\begin{equation}
\label{eq3}
c_m (\omega, z)=\frac{1}{a^2 J_1^2 ( \alpha_m )} \int_0^a \tilde A (r,z,\omega ) J_0 \left( \alpha_m \frac{r}{a} \right) r dr,
\end{equation}
being $a$ the core radius of the HCF and $\alpha_m$ the $m^{th}$-zero of $J_0 (x)$, where $J_\nu$ is the Bessel function of the first kind of order $\nu$. We solve Eq. \ref{eq3} by using the discrete Hankel Transform scheme proposed in \cite{g-sicairos04}. The complete linear propagation in the HCF is simulated by using the complex propagation coefficient of each mode, $\beta_m (\omega)$, \cite{marcatili64}, as shown in Eq. \ref{eq4}. The real and imaginary parts of $\beta_m (\omega)$ take into account all the dispersion and losses of the $m^{th}$-mode of the HCF, respectively
\begin{equation}
\label{eq4}
\tilde A (r,z+\Delta z,\omega ) = \sum_{m=1}^{\infty} {c_m (\omega,z) EH_{1m} (r) \exp (i\beta_m (\omega) \Delta z)}.
\end{equation}

The second sub-step of the method consists in applying the nonlinear effects. Separating $\hat N \left[A(r,z,T) \right]=N_{SPM} (A) + N_{ioniz} (A) + N_{abs} (A)$, a mathematical expression for each term can be obtained\cite{couairon07}, as given below. For the self-phase modulation, Raman scattering and self-steepening, we have
\begin{equation}
\label{eq6}
N_{SPM} (A)= ik_0 n_2 \left( 1+\frac{i}{\omega_0} \frac{\partial}{\partial T} \right) \left(  A(r,z,T) \int_{-\infty}^T {K(T-t) |A(r,z,T)|^2 dt} \right).
\end{equation}
In Eq. \ref{eq6} $k_0 =n_0 2\pi /\lambda_0$, with $\lambda_0$ the central wavelength of the pulse. $n_2$ the nonlinear refractive index, $\omega_0 =2\pi c / \lambda_0$ and $K(T-t)$ representing the SPM together with the Raman scattering, which has the following form $(1-x)\delta(T-t) + x/ \tau_K \exp ( -(T-t)/ \tau_K )$, where $x$ is the ratio between the SPM and the Raman effect, and $\tau_K$ fs the characteristic time for the Raman response. For the ionization, we have
\begin{equation}
\label{eq7}
N_{ioniz} (A)= -\frac{\sigma}{2} (1+i\omega_0 \tau_C ) \left( 1+\frac{i}{\omega_0} \frac{\partial}{\partial T} \right)^{-1} \left[ \rho(r,T) A(r,z,T) \right],
\end{equation}
where $\sigma$ is the cross section for the inverse Bremsstrahlung that depends on the collision time $(\tau_C)$, the critical density of the medium and the central frequency of the laser pulse \cite{couairon07}. $\rho$ represents the ionized electron density, whose evolution is governed by $\partial \rho /\partial t = W(|A|^2) (\rho_{at} - \rho)$, where $W(|A|^2)$ is the ionization rate calculated using the PPT model \cite{peremolov66} and $\rho_{at}$ is the atomic density of the medium. Finally, the absorption term is
\begin{equation}
\label{eq8}
N_{abs} (A)= -\frac{W(|A|^2)U_i}{2|A|^2} (\rho_{at} -\rho),
\end{equation}
where $U_i$ represents the ionization potential of the gas.
\newline

\noindent {\bf Experiments.}  The experiments were performed by employing 23-25 fs Fourier-transform-limited laser pulses centered at 780 nm. These pulses were generated with a 1 kHz Ti:Sapphire CPA laser system (Femtolasers FemtoPower Compact PRO HE CEP) which is part of the CLPU facility. The maximum pulse energy available was 2.5\,mJ and we adjusted it to 1\,mJ in our experiments. The laser pulses were focused by a spherical mirror (1\,m focal length) into a hollow-core fiber (HCF) with an inner diameter of 250 micron and 1 meter length. The HCF was filled with argon gas at different pressures. Output pulses were compressed and measured using the d-scan technique from Sphere Ultrafast Photonics. In the scanning SHG d-scan setup we used a double-angle chirped mirror set (Ultrafast Innovations GmbH) and two motorized BK7 wedges to induce the dispersion scan.

\section*{Acknowledgements}

We acknowledge funding from the following institutions: Junta de Castilla y Le\'on (Project SA116U13, SA046U16); MINECO (FIS2013-44174-P, FIS2015-71933-REDT, FIS2016-75652-P); Funda\c{c}\~{a}o para a Ci\^{e}ncia e a Tecnologia, Portugal, (Grants UID/NAN/50024/2013, NORTE-07-0124-FEDER-000070); Consejo Nacional de Ciencia y Tecnolog\'ia, M\'exico (CONACYT M\'exico for Postdoctoral Research Fellowships). CLPU is acknowledged for granting access to its facilities.

\nolinenumbers

%This is where your bibliography is generated. Make sure that your .bib file is actually called library.bib


\begin{thebibliography}{99}

%\bibliography{library}

\bibitem{hentschel01}  M. Hentschel, R. Kienberger, C. Spielmann, G. A. Reider, N. Milosevic, T. Brabec, P. Corkum, U. Heinzmann, M. Drescher, and F. Krausz, "Attosecond metrology," Nature {\bf 414}, 509-513 (2001).

\bibitem{kling2008}  M. F. Kling, and M.J.J. Vrakking, "Attosecond electron dynamics," Annu. Rev. Phys. Chem. {\bf 59}, 463–492  (2008).

\bibitem{krausz2009} F. Krausz, and M. Ivanov, "Attosecond physics," Rev. Mod. Phys. {\bf 81}, 163-264 (2009).

\bibitem{gallmann2012}  L. Gallmann, C. Cirelli, and U. Keller, "Attosecond science: recent highlights and future trends," Annu. Rev. Phys. Chem. {\bf 63}, 447–469  (2012).

\bibitem{krausz2014}  F. Krausz, and M. I. Stockman, "Attosecond metrology: from electron capture to future signal processing," Nat. Photonics {\bf 8}, 205-213 (2014).

\bibitem{timmers2017} H. Timmers, Y. Kobayashi, K.F. Chang, M. Reduzzi, D.M. Neumark, and S.R. Leone, "Generating high-contrast, near single-cycle waveforms with third-order dispersion compensation," Opt. Lett. {\bf 42}, 811-814 (2017).

\bibitem{sansone06} G. Sansone, E. Benedetti, F. Calegari, C. Vozzi, L. Avaldi, R. Flammini, L. Poletto, P. Villoresi, C. Altucci, R. Velotta, S. Stagira, S. De Silvestri, and M. Nisoli, "Isolated single-cycle attosecond pulses," Science {\bf 314}, 443-446 (2006).

\bibitem{polli2008} D. Polli, M.R. Antognazza, D. Brida, G. Lanzani, G. Cerullo, and S. De Silvestri, "Broadband pump-probe spectroscopy with sub-10-fs resolution for probing ultrafast internal conversion and coherent phonons in carotenoids," Chem. Phys. {\bf 350}, 45-55 (2008).

\bibitem{liebel2015} M. Liebel, C. Schnedermann, T. Wende, and P. Kukura, "Principles and applications of broadband impulsive vibrational
spectroscopy," J. Chem. Phys. A {\bf 119}, 9506-9517 (2015).

\bibitem{du2016} J. Du, J. Harra, M. Virkki, J.M. M\"akel\"a, Y. Leng, M. Kauranen, and T. Kobayashi, "Surface-enhanced impulsive coherent vibrational spectroscopy," Sci. Rep. {\bf 6}, 36471 (2016).

\bibitem{kukura2007} P. Kukura, D.W. McCamant, and R.A. Mathies, "Femtosecond stimulated Raman spectroscopy," Annu. Rev. Phys. Chem.  {\bf 58}, 461-488 (2007).

\bibitem{fujisawa2016} T. Fujisawa, H. Kuramochi, H. Hosoi, S. Takeuchi, and T. Tahara, "Role of coherent low-frequency motion in excited-state proton transfer of green fluorescent protein studied by time-resolved impulsive stimulated Raman spectroscopy," J. Am. Chem. Soc., {\bf 138}, 3942-3945 (2016).

\bibitem{dietze2016} D.R. Dietze, and R.A. Mathies, "Femtosecond stimulated Raman spectroscopy," ChemPhysChem {\bf 17}, 1224-1251 (2016).

\bibitem{kuramochi2016} H. Kuramochi, Sa. Takeuchi, and T. Tahara, "Femtosecond time-resolved impulsive stimulated Raman spectroscopy using sub-7-fs pulses: Apparatus and applications," Rev. Sci. Instrum. {\bf 87}, 043107 (2016).

\bibitem{wang2015} Y.-T. Wang, M.-H. Chen, C.-T. Lin, J.-J. Fang, C.-J. Chang, C.-W. Luo, A. Yabushita, K.-H. Wu, and T. Kobayashi, “Use of ultrafast time-resolved spectroscopy to demonstrate the effect of annealing on the performance of P3HT:PCBM solar cells,” ACS Appl. Mater. Interfaces {\bf 7}, 4457–4462 (2015).

\bibitem{luo2016} C.-W. Luo, Y.-T. Wang, A. Yabushita, and T. Kobayashi, "Ultrabroadband time-resolved spectroscopy in novel types of condensed matter," Optica {\bf 3}, 82-92 (2016).

\bibitem{schnedermann2016} C. Schnedermann, J. M. Lim, T. Wende, A.S. Duarte, L. Ni, Q. Gu, A. Sadhanala, A. Rao, and P. Kukura, "Sub-10 fs time-resolved vibronic optical microscopy," J. Phys. Chem. Lett. {\bf 7}, 4854-4859 (2016).

\bibitem{nishiyama2015} Y. Nishiyama, K. Imura, and H. Okamoto, "Observation of plasmon wave packet motions via femtosecond time-resolved near-field imaging techniques," Nano Lett. {\bf 15}, 7657-7665 (2015).

\bibitem{darmo2004} J. Darmo, T. M\''uller, W. Parz, J. Kr\''oll, G. Strasser, and K. Unterrainer, "Few-cycle terahertz generation and spectroscopy of nanostructures," Phil. Trans. R. Soc. Lond. A {\bf 362}, 251-262 (2004).

\bibitem{seres03} J. Seres, A. Muller, E. Seres, K. O'Keeffe, M. Lenner, R. Herzog, D. Kaplan, C. Spielmann, and F. Krausz, "Sub-10-fs, terawatt-scale Ti : sapphire laser system," Opt. Lett. {\bf 28}, 1832-1834 (2003).

\bibitem{cerullo98} G. Cerullo, M. Nisoli, S. Stagira, and S. De Silvestri, "Sub-8-fs pulses from an ultrabroadband optical parametric amplifier in the visible," Opt. Lett. {\bf 23}, 1283-1285 (1998).

\bibitem{miranda12b} M. Miranda, C.L. Arnold, T. Fordell, F. Silva, B. Alonso, R. Weigand, A. L'Huillier and H. Crespo, "Characterization of broadband few-cycle laser pulses with the d-scan technique," Opt. Express {\bf 20}, 18732 (2012).

\bibitem{cardin15} V. Cardin, N. Thir\'e, S. Beaulieu, V. Wanie, F. L\'egar\'e and B.E. Schmidt, "0.42 TW 2-cyle pulses at 1.8 $\mu$m via hollow-core fiber compression," Appl. Phys. Lett. {\bf 107}, 181101 (2015).

\bibitem{tomlinson84} W. J. Tomlinson, R. H. Stolen, and C. V. Shank, "Compression of optical pulses chirped by self-phase modulation in fibers," J. Opt. Soc. Am. B {\bf 1}, 139-149 (1984).

\bibitem{fork87} R. Fork, C. Cruz, P. Becker, and C. Shank, "Compression of optical pulses to 6 femtoseconds by using cubic phase compensation," Opt. Lett. {\bf 12}, 483-485 (1987).

\bibitem{nisoli96} M. Nisoli, S. DeSilvestri and O. Svelto, "Generation of high energy 10 fs pulses by a new pulse compression technique," Appl. Phys. Lett. {\bf 68}, 2793 (1996).

\bibitem{sartania97}  S. Sartania, Z. Cheng, M. Lenzner, G. Tempea, Ch. Spielmann, F. Krausz, and K. Ferencz, "Generation of 0.1-TW 5-fs optical pulses at a 1-kHz repetition rate," Opt. Lett. {\bf 22}, 1562-1564 (1997).

\bibitem{hauri04} C. P. Hauri, W. Kornelis, F. W. Helbing, A. Heinrich, A. Couairon, A. Mysyrowicz, J. Biegert, and U. Keller, "Generation of intense, carrier-envelope phase-locked few-cycle laser pulses through filamentation," Appl. Phys. B-Lasers and Optics {\bf 79}, 673-677 (2004).

\bibitem{blattermann2015} A. Bl\"attermann, C. Ott, A. Kaldun, T. Ding, V. Stoo\"u, M. Laux, M. Rebholz, and T. Pfeifer, "In situ characterization of few-cycle laser pulses in transient absorption spectroscopy," Opt. Lett. {\bf 40}, 3464-3467 (2015). 

\bibitem{schiffrin2013} A. Schiffrin, T. Paasch-Colberg, N. Karpowicz, V. Apalkov, D. Gerster, S. M\''hlbrandt, M. Korbman, J. Reichert, M. Schultze, S. Holzner, J.V. Barth, R. Kienberger, R. Ernstorfer, V.S. Yakovlev, M.I. Stockman, and F. Krausz, "Optical-field-induced current in dielectrics," Nature {\bf 493}, 70-74 (2013).

\bibitem{schultze2013} M. Schultze, E.M. Bothschafter, A. Sommer, S. Holzner, W. Schweinberger, M. Fiess, M.Hofstetter, R. Kienberger, V. Apalkov, V.S. Yakovlev, M.I. Stockman, and F. Krausz, "Controlling dielectrics with the electric field of light," Nature {\bf 493}, 75-78 (2013).

\bibitem{bocharova2011} I. A. Bocharova, A.S. Alnaser, U. Thumm, T. Niederhausen, D. Ray, C. L. Cocke, and I. V. Litvinyuk, "Time-resolved Coulomb-explosion imaging of nuclear wave-packet dynamics induced in diatomic molecules by intense few-cycle laser pulses," Phys. Rev. A {\bf 83}, 013417 (2011).

\bibitem{liu2010} J. Liu, K. Okamura, Y. Kida, T. Teramoto, and T. Kobayashi, "Clean sub-8-fs pulses at 400 nm generated by a hollow fiber compressor for ultraviolet ultrafast pump-probe spectroscopy," Opt. Express {\bf 18}, 20645-20650 (2010).

\bibitem{liu2013} J. Liu, A. Yabushita, S. Taniguchi, H. Chosrowjan, Y. Imamoto, K. Sueda, N. Miyanaga, and T. Kobayashi, "Ultrafast time-resolved pump-probe spectroscopy of PYP by a sub-8 fs pulse laser at 400 nm," J. Phys. Chem. B {\bf 117}, 4818-4826 (2013).

\bibitem{kobayashi2012} T. Kobayashi, and Y. Kidaw, "Ultrafast spectroscopy with sub-10 fs deep-ultraviolet pulses," Phys. Chem. Chem. Phys. {\bf 14},  6200-6210 (2012).

\bibitem{gueye2016} M. Gueye, J. Nillon, O. Cregut, and J. L\'eonard, "Broadband UV-Vis vibrational coherence spectrometer based on a hollow fiber compressor," Rev. Sci. Instrum. {\bf 87}, 093109 (2016).

\bibitem{paolino2016} M. Paolino, M. Gueye, E. Pieri, M. Manathunga, S. Fusi, A. Cappelli, L. Latterini, D. Pannacci, M. Filatov, J. L\'eonard, and M. Olivucci, "Design, synthesis, and dynamics of a green fluorescent protein fluorophore mimic with an ultrafast switching function," J. Am. Chem. Soc., {\bf 138}, 9807–9825 (2016).

\bibitem{goncalves2016} C.S. Gonçalves, A.S. Silva, D. Navas, M. Miranda, F. Silva, H. Crespo, and D.S. Schmool, “A dual-colour architecture for pump-probe spectroscopy of ultrafast magnetization dynamics in the sub-10-femtosecond range",  Sci. Rep. {\bf 6}, 22872 (2016).

\bibitem{chang16} H.-T. Chang, M. Z\"urch, P.M. Kraus, L-J. Borja, D.M. Neumark and S.R. Leone, "Simultaneous generation of sub-5-femtosecond 400 nm and 800 nm pulses for attosecond extreme ultraviolet pump-probe spectroscopy", Opt. Lett. {\bf 41}, 5365-5368 (2016).

\bibitem{guenot16} D. Gu\'enot, D. Gustas, A. Vernier, B. Beaurepaire, F. B\"ohle, M. Bocoum, M. Lozano, A. Jullien, R. Lopez-Martens, A. Lifschitz and J. Faure, "Relativistic electron beams driven by kHz single-cycle light pulses" Nat. Photonics {\bf 11}, 293-296 (2017).

\bibitem{miranda12} M. Miranda, T. Fordell, C. Arnold, A. L'Huillier, and H. Crespo, "Simultaneous compression and characterization of ultrashort laser pulses using chirped mirrors and glass wedges," Opt. Express {\bf 20}, 688-697 (2012).

\bibitem{louisy15}  M. Louisy, C.L. Arnold, M. Miranda, E.W. Larsen, S.N. Bengtsson, D. Kroon, M. Kotur, D. Gu\'enot, L. Rading, P. Rudawski, F. Brizuela, F. Campi, B. Kim, A. Jarnac, A. Houard, J. Mauritsson, P. Johnsson, A. L'Huillier and C.M. Heyl, "Gating attosecond pulses in a noncollinear geometry," Optica {\bf 2} 563-566 (2015) (Supplementary Material).

\bibitem{miranda16} M. Miranda, J. Penedones, C. Guo, A. Harth, M. Louisy, L. Neoricic, A. L'Huillier, and C. L. Arnold, "Fast iterative retrieval algorithm for ultrashort pulse characterization using dispersion scan," J. Opt. Soc. Am. B {\bf 34}, 1190-197 (2017).

\bibitem{silva14} F. Silva, M. Miranda, B. Alonso, J. Rauschenberger, V. Pervak, and H. Crespo, "Simultaneous compression, characterization and phase stabilization of GW-level 1.4 cycle VIS-NIR femtosecond pulses using a single dispersion-scan setup," Opt. Express {\bf 22}, 10181-10190 (2014).

\bibitem{fabris15} D. Fabris, W. Holgado, F. Silva, T. Witting, J.W.G. Tisch and H. Crespo, "Single-shot implementation of dispersion-scan for the characterization of ultrashort laser pulses," Opt. Express {\bf 23}, 32803-32808 (2015).

\bibitem{fan16} G. Fan, T. Balciunas, T. Kanai, T. Fl\"ory, G. Anriukaitis, B.E. Schmidt, F. L\'egar\'e and A. Baltuska, "Hollow-core-waveguide compression of multi-millijoule CEP-stable 3.2 $\mu$m pulses," Optica {\bf 3}, 1308-1311 (2016).

\bibitem{Alonso13} B. Alonso, M. Miranda, F. Silva, V. Pervak, J. Rauschenberger, J. San Roman, I.J. Sola, and H. Crespo, "Characterization of sub-two-cycle pulses from a hollow-core fiber compressor in the spatiotemporal and spatiospectral domains," Appl. Phys. B {\bf 112}, 105-114 (2013).

\bibitem{bohle14} F. Bohle, M. Kretschmar, A. Jullien, M. Kovacs, M. Miranda, R. Romero, H. Crespo, U. Morgner, P. Simon, R. Lopez-Martens, and T. Nagy, "Compression of CEP-stable multi-mJ laser pulses down to 4 fs in long hollow fibers," Las. Phys. Lett. {\bf 11} 095401 (2014).

\bibitem{heyl16} C. M. Heyl, H. Coudert-Alteirac, M. Miranda, M. Louisy, K. Kovacs, V. Tosa, E. Balogh, K. Varj\'u, A. L'Huillier, A. Couairon, and C. L. Arnold, "Scale-invariant nonlinear optics in gases," Optica {\bf 3}, 75-81 (2016) (Supplementary Material).

\bibitem{tajalli16} A. Tajalli, B. Chanteau, M. Kretschmar, H.G. Kurz, D. Zuber, M. Kovacev, U. Morgner and T. Nagy, "Few-cycle optical pulse characterization via cross-polarized wave generation dispersion scan technique," Opt. Lett. {\bf 41}, 5246-5249 (2016).

\bibitem{schmidt10} B. E. Schmidt, et al. "Compression of 1.8 $\mu$m laser pulses to sub two optical cycles with bulk material," Appl. Phys. Lett. {\bf 96}, 121109 (2010).

\bibitem{huang16} Z. Huang, D. Wang, Y. Dai, Y. Li, X. Guo, W. Li, Y. Chen, J. Lu, Z. Liu, R. Zhao and Y. Leng, "Design of intense 1.5-cycle pulses generation at 3.6 $\mu$m through a pressure gradient hollow-core fiber," Opt. Express {\bf 24}, 9280-9287 (2016). 


\bibitem{marcatili64} E. Marcatili and R. Schmeltzer, "Hollow metallic and dielectric waveguides for long distance optical
transmission and lasers," Bell Syst. Tech. J. {\bf 43}, 1783-1809 (1964).

\bibitem{couairon07} A. Couairon, and A. Mysyrowicz, "Femtosecond filamentation in transparent media," Phys. Rep. {\bf 441}, 47-189 (2007).

\bibitem{agrawal} G.P. Agrawal, "Nonlinear Fiber Optics," 3erd Ed. (Academic Press, San Diego, 2001).

\bibitem{granados12} E. Granados, L. Chen, C. Lai, K. Hong, and F. Kartner, "Wavelength scaling of optimal hollow-core fiber compressors in the single-cycle limit," Opt. Express {\bf 20}, 9099-9108 (2012).

\bibitem{g-sicairos04} M. Guizar-Sicairos and J.C. Gutierrez-Vega, "Computation of quasi-discrete Hankel transforms of integer order for propagating optical wave fields," J. Opt. Soc. Am. A {\bf 21}, 53 (2004).

\bibitem{peremolov66} A.M. Peremolov, V.S. Popov and M.V. Terentev, "Ionization of atoms in an alternating electric field," Sov. Phys. JETP {\bf 23}, 924 (1966).

\end{thebibliography}
\end{document}